\documentclass[%
10pt,A4paper,twocolumn,aps,pra, superscriptaddress,longbibliography, nofootinbib,
 amsmath,amssymb,
]{revtex4-2}
\usepackage{lipsum,mathptmx,etoolbox}
\usepackage{mathtools}

\usepackage{graphicx}

\usepackage{dcolumn}
\usepackage{bm}
\usepackage{xcolor}
\usepackage{siunitx}
\usepackage{booktabs} 
\usepackage[colorlinks=true, linkcolor=blue, citecolor=black, urlcolor=blue, pdfauthor={Yongxin Song}, pdftitle={Constant-Depth Fan-Out with Real-Time Feedforward on a Superconducting Quantum Processor}]{hyperref}

\usepackage{makecell}
\usepackage{enumitem}
\setlist[itemize]{leftmargin=*}

\usepackage[shortcuts]{extdash}

\begin{document}
\renewcommand{\arraystretch}{1.1} 
\preprint{APS/123-QED}

\title{Realization of Constant-Depth Fan-Out with Real-Time Feedforward \\on a Superconducting Quantum Processor}

\author{Yongxin Song}
\author{Liberto Beltr\'an}
\author{Ilya Besedin}
\author{Michael Kerschbaum}
\author{Marek Pechal}
\affiliation{Department of Physics, ETH Zurich, CH-8093 Zurich, Switzerland}
\affiliation{Quantum Center, ETH Zurich, CH-8093 Zurich, Switzerland}
\affiliation{ETH Zurich - PSI Quantum Computing Hub, Paul Scherrer Institute, CH-5232 Villigen, Switzerland}

\author{François Swiadek}
\author{Christoph Hellings}
\author{Dante Colao Zanuz}
\affiliation{Department of Physics, ETH Zurich, CH-8093 Zurich, Switzerland}
\affiliation{Quantum Center, ETH Zurich, CH-8093 Zurich, Switzerland}

\author{Alexander Flasby}
\affiliation{Department of Physics, ETH Zurich, CH-8093 Zurich, Switzerland}
\affiliation{Quantum Center, ETH Zurich, CH-8093 Zurich, Switzerland}
\affiliation{ETH Zurich - PSI Quantum Computing Hub, Paul Scherrer Institute, CH-5232 Villigen, Switzerland}

\author{Jean-Claude Besse}
\author{Andreas Wallraff}
\affiliation{Department of Physics, ETH Zurich, CH-8093 Zurich, Switzerland}
\affiliation{Quantum Center, ETH Zurich, CH-8093 Zurich, Switzerland}
\affiliation{ETH Zurich - PSI Quantum Computing Hub, Paul Scherrer Institute, CH-5232 Villigen, Switzerland}

\begin{abstract}
When using unitary gate sequences, the growth in depth of many quantum circuits with output size poses significant obstacles to practical quantum computation. The quantum fan-out operation, which reduces the circuit depth of quantum algorithms such as the quantum Fourier transform and Shor's algorithm, is an example that can be realized in constant depth independent of the output size. Here, we demonstrate a quantum fan-out gate with real-time feedforward on up to four output qubits using a superconducting quantum processor. By performing quantum state tomography on the output states, we benchmark our gate with input states spanning the entire Bloch sphere.
We decompose the output-state error into a set of independently characterized error contributions. We extrapolate our constant-depth circuit to offer a scaling advantage compared to the unitary fan-out sequence beyond 25 output qubits with feedforward control, or beyond 17 output qubits if the classical feedforward latency is negligible. Our work highlights the potential of mid-circuit measurements combined with real-time conditional operations to improve the efficiency of complex quantum algorithms.
\end{abstract}

\date{\today}

\maketitle

\section{Introduction}
Using the outcome of mid-circuit measurements to conditionally apply quantum gates to a subset of qubits in a quantum circuit is often referred to as adaptive control. Adaptive control enables quantum circuits which are more efficient than static ones, or even ones which are otherwise unattainable~\cite{FossFeig2023}. Early experiments showed the important role of measurement-based adaptive control in applications including qubit initialization~\cite{Riste2012b, Salathe2018}, quantum teleportation~\cite{Bouwmeester1997, Barrett2004, Krauter2013, Steffen2013, Pfaff2014, Chou2018}, and state stabilization~\cite{Ofek2016, Negnevitsky2018, Andersen2019, Riste2020}. Entering the NISQ era, adaptive circuits raise interest in their potential to reduce the circuit depth of quantum algorithms~\cite{Pham2013, Piroli2021, Buhrman2023}. Compared to unitary sequences, protocols with adaptive circuits could achieve lower circuit depths~\cite{FossFeig2023} and show favorable error scaling in tasks such as CNOT gate teleportation and GHZ state preparation~\cite{Baeumer2023}. 

Constant-depth quantum fan-out is a powerful protocol enabled by adaptive control~\cite{Hoyer2004, Pham2013}. Quantum fan-out increases the parallelization of instructions and allows the efficient implementation of quantum computing tasks, such as constant-depth quantum Fourier transform~\cite{Hoyer2004}, constant-depth extraction of Hamming weights~\cite{Hoyer2004}, and polylogarithmic-depth Shor's algorithm~\cite{Pham2013}. Recently, a 1-to-3 quantum fan-out gate was demonstrated with superconducting qubits~\cite{Hashim2024}. The fan-out operation was applied to the computational basis states, and the performance was limited by measurement-induced cross-dephasing. 

Here, we present a constant-depth quantum fan-out gate with real-time feedforward and up to 4 output qubits. 
Adding to the above experiment, we realize quantum fan-out on input states parameterized over the Bloch sphere, and thoroughly decompose the error of the sequence.
In Section~\ref{sec:circuit_and_implementation}, we introduce the quantum fan-out gate sequence and describe its implementation on a 17-qubit superconducting quantum processor~\cite{Krinner2022}. In Section~\ref{sec:fan_out_gate_characterization}, we present our main results. We start by examining the density matrices and the multi-qubit Pauli operator expectation values of two example output states, showing that the fan-out gate yields the expected output. Then, we prepare the input state in a set of states parameterized by polar and azimuthal angles on the Bloch sphere and analyze the output states with quantum state tomography. Furthermore, we investigate the scaling of the fan-out gate error with the number of output qubits and decompose the measured error into individually characterized error sources. To understand the gate performance in the absence of the feedforward delay set by the latency of the control electronics, we replace the physical application of a feedforward control pulse with a rotation of the Pauli reference frame in post-measurement analysis. We observe a reduction in the measured error in agreement with the contribution expected from qubit decoherence during the idling time required to instantiate the feedforward pulses. Finally, we compare the performance of the constant-depth fan-out circuit with the static, unitary fan-out circuit. Extrapolating the error scaling suggests that the constant-depth circuit yields a benefit beyond 25 output qubits with feedforward control, or beyond 17 output qubits if the electronics latency can be reduced to negligible levels.

\section{Circuit and implementation} \label{sec:circuit_and_implementation}

\begin{figure}[t!]
\includegraphics{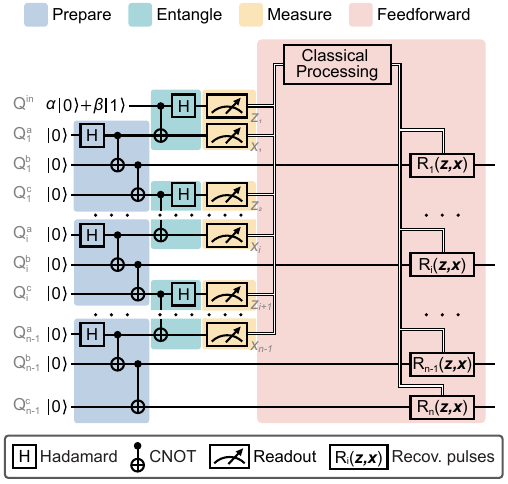}
\caption{\label{fig:circuit_diagram} Circuit diagram of the constant-depth quantum fan-out gate. Background colors mark the different steps of the circuit.}
\end{figure}

We implement a constant-depth quantum fan-out gate with $n$ output qubits based on a teleportation-like protocol with GHZ states~\cite{Pham2013}. The protocol requires $3n-2$ qubits arranged in a linear array with nearest-neighbor coupling. The first qubit in the array $Q^\mathrm{in}$ carries the input state, and the rest of the qubits can be divided into $(n-1)$ groups of three qubits $\{Q^\mathrm{a}_i, Q^\mathrm{b}_i, Q^\mathrm{c}_i\}$, see Fig.~\ref{fig:circuit_diagram}.
The protocol is executed in four time steps. 
In the first step, we simultaneously {\bf prepare} $Q^\mathrm{in}$ in the state $\alpha|0\rangle + \beta |1\rangle$ and the three qubits in each group in the GHZ state.
In the second step, we execute controlled-NOT (CNOT) gates followed by a Hadamard gate to {\bf entangle} both the input qubit and its neighboring qubit, as well as qubit pairs across the boundary of qubit groups for Bell measurements.
In the third step, we {\bf measure} all pairs of the qubits entangled in the previous step, yielding the measurement results in the computational basis $(z_1, x_1, ..., z_{n-1}, x_{n-1}) \equiv (\bm{z}, \bm{x})$, where $z_i$ and $x_i$ are the results of the first and second qubit in each Bell basis measurement.
In the fourth step, we use classical control instruments to {\bf feedforward} recovery pulses $R_q(\bm{z}, \bm{x})$ to the qubits. $R_q(\bm{z}, \bm{x})$ is a product of conditional X and Z gates, where the application of the X gate depends on the parity of all x-type qubit results $\{x_m\}$ for qubit indices smaller than or equal to the output qubit considered, i.e.~$m\leq q$, and the application of the Z gate is conditioned on the result of the closest z-type qubit result $z_q$. An exception applies to the last recovery pulse $R_n(\bm{z}, \bm{x})$, where the conditional Z gate is not needed. $R_q(\bm{z}, \bm{x})$ can be expressed as
\begin{equation} \label{eq:recover_pulse_rules}
    R_q(\bm{z}, \bm{x})= 
    \begin{cases} \begin{array}{ll}
        Z^{z_q} \prod_{m=1}^q X^{x_m},\, & 1\leq q \leq n-1 \\
        \prod_{m=1}^{n-1} X^{x_m}, \,\,\,\,\,\,\,\,\,\, & q=n.
    \end{array} \end{cases}
\end{equation}
In the absence of errors, the output qubits are deterministically prepared in the target state $\alpha|0\cdots 0\rangle + \beta |1\cdots 1\rangle$.

We apply the gate sequence to a one-dimensional chain of ten transmon qubits of a 17-qubit superconducting quantum processor~\cite{Krinner2022} (see sample and setup details in App.~\ref{app:sample}). With ten qubits, we realize a quantum fan-out with four output qubits. The control sequence is decomposed into single-qubit DRAG pulses~\cite{Motzoi2009}, virtual Z rotations~\cite{McKay2017}, and flux-activated net-zero controlled-Z (CZ) gates~\cite{Negirneac2021} (see gate sequence and time budget in App.~\ref{app:sequence}). The simultaneous readout is realized using frequency-multiplexed readout tones selectively applied to the readout resonators of the target qubits~\cite{Heinsoo2018}. The mid-circuit readout response is integrated and classified in real-time with an FPGA embedded in each control electronics, and the readout result of each qubit, as a single-bit value, is forwarded to a central control unit. The set of all readout results is interpreted with programmable lookup tables to one 2-bit index per output qubit, uniquely defining the feedforward pulse to be applied. The indices are forwarded to the drive instruments of the output qubits, which apply the single-qubit rotation pulses accordingly. The whole sequence lasts $\SI{1888}{ns}$ independent of the output qubit number, with a $\SI{800}{ns}$ latency required of the classical processing for the feedforward control (see the breakdown of time budget in App.~\ref{app:sample} and \ref{app:sequence}).

\section{Fan-out Gate characterization} \label{sec:fan_out_gate_characterization}
\begin{figure}[t!]
\includegraphics{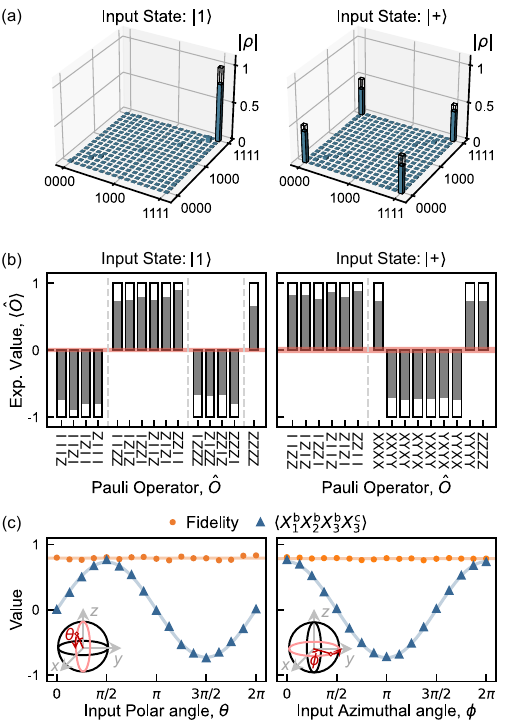}
\caption{\label{fig:different_perp_angles} (a) The absolute value of the maximum-likelihood estimation of density matrix $\rho$ when the input qubit is prepared in the $|1\rangle$ (left) and $|+\rangle$ (right) state. Black frames indicate the expected outcome for an ideal fan-out circuit. (b) Measured (gray bars) and ideal (black frame)  expectation values of the multi-qubit Pauli operators $\hat O$ of the output state with non-zero ideal values when the input is prepared in the $|1\rangle$ (left) and $|+\rangle$ (right) state. The top and bottom of the red-shaded area indicate the maximum and minimum of the measured expectation values of the rest of the Pauli operators whose ideal value is zero. (c) Extracted output state fidelity (orange circles) and the expectation values of the joint-X operators of the output qubits (blue triangles) for input states parameterized by the polar angle $\theta$ (fixing the azimuthal angle $\phi=0$) and $\phi$ (fixing $\theta=\pi/2$). Solid lines indicate the average output state fidelity (orange) and fits to a sinusoidal function (blue).}
\end{figure}

We execute the feedforward-based 1-to-4 fan-out gate with input state $|1\rangle$ and characterize the output state using quantum state tomography~\cite{Steffen2006a}. We compare $\rho$ to the ideal density matrix $\rho_\mathrm{ideal}$ (black wireframes) and extract an experimental output state fidelity $F=(\mathrm{Tr}\sqrt{ \sqrt{\rho}\rho_\mathrm{ideal}\sqrt{\rho}})^2=0.797$, see density matrix $\rho$ obtained using maximum-likelihood estimation in Fig.~\ref{fig:different_perp_angles}a. Similarly, we execute the experimental sequence using the equal superposition state $|+\rangle$ as the input and extract an output state fidelity of $0.803$, see Fig.~\ref{fig:different_perp_angles}a. 

In both reconstructed density matrices, we do not observe obvious unexpected off-diagonal elements, suggesting that coherent errors are small. We further verify this by calculating the expectation value of multi-qubit Pauli operators from the density matrices. From the 4-qubit output states of the input state $|1\rangle$, we extract the measured expectations values of Pauli operators with non-zero ideal values and observe measured values on average 0.76(7) times their ideal value of $\pm 1$ (see Fig~\ref{fig:different_perp_angles}b). In contrast, the measured expectation values that ideally vanish remain between $\pm0.05$. We observe the same fractional errors for the $|+\rangle$ input state.

To characterize our 1-to-4 fan-out gate acting on arbitrary input states, we prepare $Q^\mathrm{in}$ in a family of superposition states $|\psi_\mathrm{in}\rangle=\cos(\theta/2)|0\rangle+e^{i\phi}\sin(\theta/2)|1\rangle$ parametrized by the polar angle $\theta$ and the azimuthal angle $\phi$. For each prepared state, we characterize the corresponding output state using quantum state tomography. First, we sweep the polar angle $\theta$ while fixing the azimuthal angle $\phi=0$, essentially varying the excitation probability in the superposition state. We find an average fidelity of $0.79(2)$, see Fig.~\ref{fig:different_perp_angles}c (orange). 
As varying $\theta$ changes the X-parity of the output state, we calculate $\langle X_1^\mathrm{b} X_2^\mathrm{b} X_3^\mathrm{b} X_3^\mathrm{c}\rangle$ from the reconstructed density matrices (blue) and observe that it follows a sinusoidal oscillation in the expectation value of the joint-X operator as expected for the ideal state. 
We fit the contrast of the sine oscillation to $0.730(6)$, with the ideal value being 1. The measured contrast being $92\%$ of the average fidelity is a consequence of dephasing errors and mid-circuit readout errors, which affect the joint-X operator more strongly than the fidelity. Similarly, we fix the polar angle $\theta =\pi/2$ and sweep the azimuthal angle $\phi$, varying the phase of an equal superposition state. The sweep gives an average fidelity of $0.78(1)$ and a joint-X operator oscillation contrast of $0.736(4)$.
As observed, the fidelity is primarily independent of the input state. This suggests that the feedforward-based fan-out sequence which we have realized performs uniformly well for arbitrary input states and has no prominent coherent errors.

\begin{figure}[t!]
\includegraphics{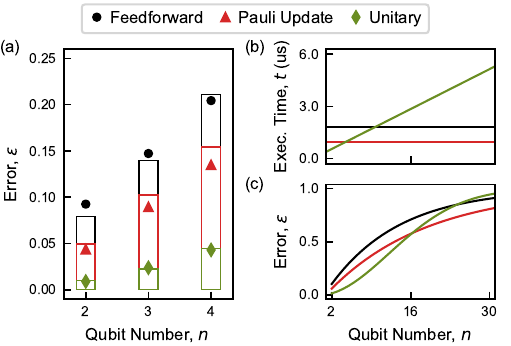}
\caption{\label{fig:distance_scaling} (a) Fan-out gate error $\varepsilon$ vs.~output qubit number $n$ with feedforward control (black circles), Pauli-frame update in post-processing (red triangles), and the unitary gate sequence (green diamonds). Wireframes of the same color show the expected error derived from a success probability model, taking into account individual CNOT gate errors, readout errors, and idling errors.  (b) Circuit execution time of the three implementations. (c) Expected error calculating using a success probability model.}
\end{figure}

Furthermore, we study the scaling of the output state error $\varepsilon = 1-F$ with the number $n$ of output qubits. We prepare the input state in the six cardinal states $|0/1\rangle,|\pm\rangle,|\pm i\rangle$ and extract an average error of $0.09(1), 0.14(1), 0.20(2)$ for $n=2, 3,$ and $4$, respectively. We compare the measured errors with a success probability model that considers CNOT gate errors, readout errors, and idling errors, assuming that the errors are uncorrelated, see App.~\ref{app:error_scaling}. We observe quantitative agreement between the measured and the expected error, see Fig~\ref{fig:distance_scaling}a. 

To investigate the limit of the expected fan-out gate performance independent of the feedforward latency, we run a constant-depth fan-out sequence with Pauli-frame update~\cite{Knill2005}. To do so, we read out the output qubits when we perform the mid-circuit measurement and update the Pauli reference frame in post-processing. 
In this scenario, we avoid the contribution to the total error stemming from qubit idling during the feedforward part of the sequence, while all other error sources remain. For the 2\=/, 3\=/, and 4\=/qubit output states, we obtain an average error of $0.044(3)$, $0.09(1)$, and $0.14(2)$ over the six cardinal input states. For comparison, we calculate the error expected from the success probability model, excluding the idling error introduced during the feedforward latency, and observe quantitative agreement between the measured and expected error. Our success probability model being verified by both the feedforward and the Pauli-frame update data supports our assumption that the errors from different sources are uncorrelated and also allows us to analyze the ideal case in which the latency from the electronics is negligible.

For reference, we run a static, unitary fan-out circuit between nearest-neighbor qubits, see App.~\ref{app:sequence}, and compare its performance with the constant-depth (i.e.~feedforward and Pauli-frame update) protocols.
At two to four output qubits, the executed unitary circuits show smaller errors than the equivalent constant-depth circuits, see Fig.~\ref{fig:distance_scaling}a. This is expected from the shallower circuit depth, fewer gates, and the smaller number of qubits in the unitary protocol. However, we expect the constant-depth circuits to perform better with a larger number of output qubits. Indeed, the execution time increases linearly with the output qubit number $n$ in the case of the unitary circuit, while it remains fixed for the constant-depth circuit, see Fig.~\ref{fig:distance_scaling}b. The different scaling of the execution time causes the idling error to scale quadratically for the unitary circuit and linearly for the constant-depth circuits, see App.~\ref{app:error_scaling}, which leads to the benefit of the constant-depth circuit at large output qubit numbers. We calculate the expected output state error of the implemented protocols with our success probability model.
We find that the feedforward implementation is expected to outperform the unitary implementation when the fan-out is applied to 25 or more output qubits, while the Pauli frame update implementation shows an advantage over the unitary circuit for 17 output qubits or more, see Fig.~\ref{fig:distance_scaling}c.

\section{Discussion and Outlook}
In this work, we present a prototypical 1-to-4 fan-out gate with real-time signal processing and feedforward control. We characterize the output states with quantum state tomography and extract an average fidelity of 0.79(2) and 0.78(1) for the input states parameterized over the polar and azimuthal angles.
We study the scaling of the output state error, which we decompose into individually characterized error sources, with the output qubit number. Compared with the unitary circuit, the constant-depth protocol is expected to yield a benefit at 25 output qubits with the real-time feedforward control realized in this experiment or at 17 output qubits in the limit at which the feedforward delay is negligible. 

Our work shows that the performance of a fan-out gate on a linear nearest-neighbor qubit array can be optimized by reducing the circuit depth with adaptive circuits. Further improvements could be made by reducing the sequence length through faster quantum or classical operations, increasing the qubit coherence, and investigating efficient dynamical decoupling sequences during idling times. 
We emphasize that we implement the circuit with real-time feedforward control, which allows the output state to be used for further processing in more complex protocols, such as the quantum Fourier transform~\cite{Hoyer2004} or Shor's algorithm on a nearest-neighbor connected lattice of qubits~\cite{Pham2013}. It can also find applications in protocols requiring non-local entanglement, for example, quantum low-density parity check (LDPC) codes~\cite{Breuckmann2021}.

We note that in parallel to our work, another constant-depth quantum fan-out gate was demonstrated on superconducting circuits with up to 50 output qubits with Pauli-frame update~\cite{Baeumer2024}. The experiment observes a crossover at 7 qubits, above which the constant-depth circuit outperforms the unitary circuit. We find the result similar to our work. The two works complement each other, as Ref.~\cite{Baeumer2024} experimentally demonstrates an advantage of the constant-depth circuit by extending the system size, while we implement the real-time feedforward recovery pulses required for using the output states in a quantum algorithm and study the error composition.

\section*{Acknowledgments}
The authors thank Niels Neumann, Marten Folkertsma, and Harry Buhrman for insightful discussions.

The authors acknowledge financial support by the Swiss State Secretariat for Education, Research and Innovation (SERI) under contract number UeM019-11, by Innosuisse via the Innovation project (104.020 IP-ICT / Agreement Nr.~2155012229), 
by the Intelligence Advanced Research Projects Activity (IARPA) and the Army Research Office, under the Entangled Logical Qubits program and Cooperative Agreement Number W911NF-23-2-0212,
by the SNSF R'equip grant 206021-170731, by the Baugarten Foundation and the ETH Zurich Foundation, and by ETH Zurich. The views and conclusions contained in this document are those of the authors and should not be interpreted as representing the official policies, either expressed or implied, of IARPA, the Army Research Office, or the U.S. Government. The U.S. Government is authorized to reproduce and distribute reprints for Government purposes notwithstanding any copyright notation herein.

\section*{Author contribution}
Y.\,S. and J.-C.\,B. conceived the project. Y.\,S. acquired the data. Y.\,S. and J.-C.\,B. contributed to the analysis. Y.\,S., L.\,B. and C.\,H. contributed to the classical feedforward development. I.\,B., M.\,K. and M.\,P. contributed to the measurement setups. D.\,C.\,Z. and A.\,F. fabricated the device designed by F.\,S.. J.-C.\,B. and A.\,W. supervised the project. Y.\,S. and \mbox{J.-C.\,B}. wrote the manuscript with inputs from all authors.

\appendix
\section{Sample and Experimental Setup}

\label{app:sample}

\begin{figure}[t!]
\includegraphics{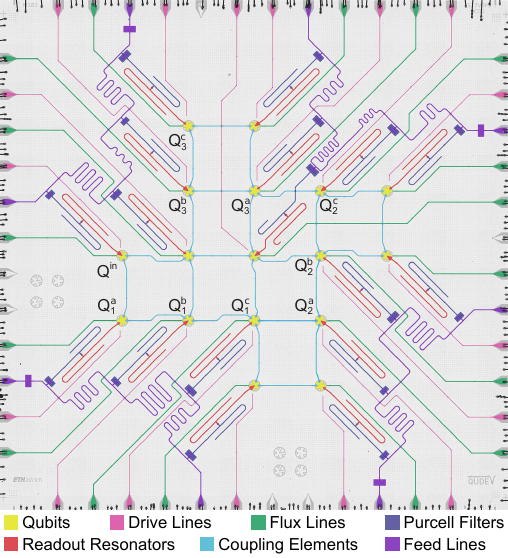}
\caption{\label{fig:device_image} False-color micrograph of the device used for the experiment. The qubits involved in the fan-out gate are labeled in the figure.}
\end{figure}

We perform the experiment on the 17-qubit superconducting quantum processor shown in Fig.~\ref{fig:device_image}. The transmon islands, coplanar waveguides, and couplers are patterned with photolithography and reactive-ion etching into a 150-nm niobium thin film sputtered onto a high-resistivity silicon substrate. Aluminium-titanium-aluminium trilayer airbridges connect the otherwise separated ground planes. We fabricate the aluminium-based Josephson junctions of the transmon qubits using electron-beam lithography (EBL) and shadow evaporation.

\begin{figure}[t!]
\includegraphics{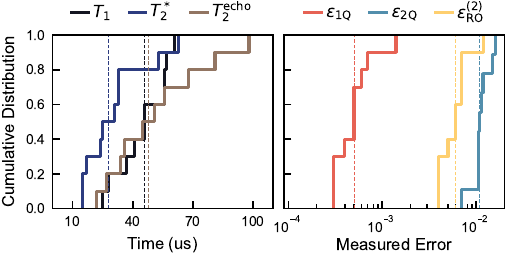}
\caption{\label{fig:cumulative_error_distribution} Cumulative distributions (integrated histograms) of (left panel) qubit relaxation time $T_1$ (black), dephasing time $T_2^*$ (blue), and Hahn-echo dephasing time $T_2^\mathrm{echo}$ (brown) ; (right panel) single-qubit gate (red), simultaneous two-qubit gate (cyan), and two-state readout (yellow) errors. Vertical dashed lines indicate the median values of the corresponding quantity.}
\end{figure}

We characterize the coherence property of each qubit with standard methods~\cite{Schreier2008}, see Fig.~\ref{fig:cumulative_error_distribution}. We extract a median relaxation time $T_1=46\,\mu\mathrm{s}$, dephasing time $T_2^*=28\,\mu\mathrm{s}$ and Hahn-echo dephasing time $T_2^\mathrm{echo}=48\,\mu\mathrm{s}$.

In the experiment, we implement single-qubit gates with 32-ns microwave DRAG pulses~\cite{Motzoi2009} with Gaussian envelopes truncated at $\pm 2.5\sigma$, where $\sigma=\SI{6.4}{ns}$. The CZ gates are realized with net-zero flux pulses~\cite{Rol2019, Negirneac2021} with an average duration of $\SI{104}{ns}$. We add $\SI{20}{ns}$ buffers before and after each flux pulse to account for the finite size of the pre-distortion filters~\cite{Rol2020} and round each total CZ gate time to an integer multiple of $\SI{8}{ns}$ to be commensurate with the granularity of the waveform playback of the control instruments. This results in an average total duration of the CZ gate of $\SI{144}{ns}$.

We benchmark the performance of the single-qubit gates with randomized benchmarking~\cite{Magesan2011, Epstein2014} and the performance of two-qubit gates with interleaved randomized benchmarking~\cite{Magesan2012, Corcoles2013, Barends2014}, see Fig.~\ref{fig:cumulative_error_distribution}b. We find a median single-qubit error $\varepsilon_\mathrm{1Q}=0.05\%$ and two-qubit error $\varepsilon_\mathrm{2Q}=1.1\%$. We characterize the single-shot readout performance by preparing the qubits in either computational basis state $|0\rangle$ and $|1\rangle$ and determining the readout assignment probability matrix with single-shot readout. To initialize our qubit register, we perform a qubit readout before the start of the gate sequence and reject all instances where qubits are found in the excited state. The readout error $\varepsilon_\mathrm{RO}^{(2)}$ is calculated as the average of the misassignment probability $P_{0|1}$ and $P_{1|0}$, where $P_{a|b}$ stands for the probability of reading out the qubit in state $a$ when preparing in state $b$. We find a median readout error $\varepsilon_\mathrm{RO}^{(2)}=0.6\%$.

\begin{figure}[t!]
\includegraphics{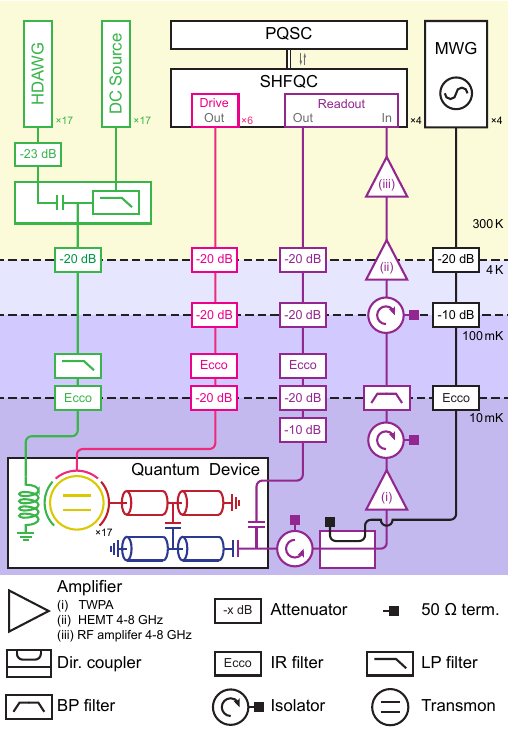}
\caption{\label{fig:cabling_diagram} Schematic diagram of the experimental setup. Flux lines (green), drive lines (pink), and readout lines (purple) transmit the signal between the control electronics and the device, including transmon qubits (yellow), readout resonators (red), and Purcell filters (blue). The background colors indicate the temperature stages of the experimental setup.}
\end{figure}

The 17-qubit device is installed at the $\SI{10}{mK}$ temperature stage of a dilution refrigerator in a standard wiring and shielding configuration~\cite{Krinner2019}. We connect the room-temperature electronics to the device with signal lines comprising microwave components and coaxial cables, see Fig.~\ref{fig:cabling_diagram}. The DC source provides a constant bias current for each qubit to set its idle operation frequency. The arbitrary waveform generators (HDAWG) generate voltage pulses at a rate of $\SI{2.0}{GSa/s}$ to implement two-qubit gates. The DC and IF signals are combined with bias-tees at room temperature. We use super-high-frequency qubit controllers (SHFQC) to drive and read out the qubits. The SHFQC drive channels generate drive pulses at the qubit transition frequencies with a double frequency conversion 
scheme~\cite{Herrmann2022a}. The SHFQC readout output channel generates multiplexed readout tones at the readout resonator frequencies, which after interacting with the device pass through an amplification chain consisting of a wide-bandwidth near-quantum limited traveling-wave parametric amplifier (TWPA)~\cite{Macklin2015}, a high-electron-mobility transistor
(HEMT) amplifier, and a low-noise, room-temperature amplifier. The amplified signal is digitized and integrated for $\SI{400}{ns}$ by the SHFQC readout input channel. After the mid-circuit measurement, the integration result is compared to a predefined threshold to discriminate the projected state of a qubit, and all thresholded results are forwarded to a programmable quantum system controller (PQSC) to be interpreted with
look-up tables as indices referring to the feedforward control pulses. The indices are subsequently sent to the SHFQC drive channels, which play the recovery pulses. The feedforward idling time is $\SI{800}{ns}$, including the signal propagation delay, the classical processing time, and a $\SI{40}{ns}$ buffer time. This control electronics architecture allows for a straightforward extension to a 60-qubit setup while maintaining the same feedforward idling time.

\section{Gate Sequence and Time Budget}\label{app:sequence}

\begin{figure}[t!]
\includegraphics{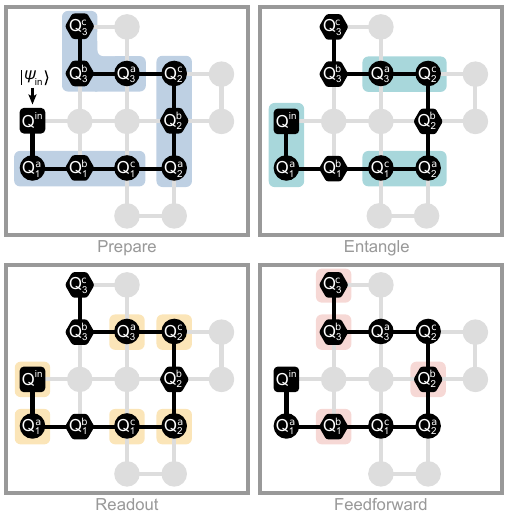}
\caption{\label{fig:gate_sequence} Mapping of a 1-to-4 fan-out gate onto a 17-qubit superconducting quantum device. Qubits involved in the algorithm are highlighted in black. $Q^\mathrm{in}$ (square) carries the input state $|\psi_\mathrm{in}\rangle$. In each step, qubits being driven or read out are highlighted with a background color. The protocol maps the four-qubit output state onto $Q_1^\mathrm{b}, Q_2^\mathrm{b}, Q_3^\mathrm{b}$ and $Q_3^\mathrm{c}$ (hexagons).}
\end{figure}

We map the gate sequence in Sec.~\ref{sec:circuit_and_implementation} to ten qubits selected from the 17-qubit quantum device, see Fig.~\ref{fig:gate_sequence}. The mapping allows for the execution of every parallel two-qubit gate step without non-interacting qubits crossing each other in frequency. Adjusting the output qubit number could be achieved by adding or removing groups of three qubits at the input qubit end.

\begin{figure*}[t!]
    \includegraphics{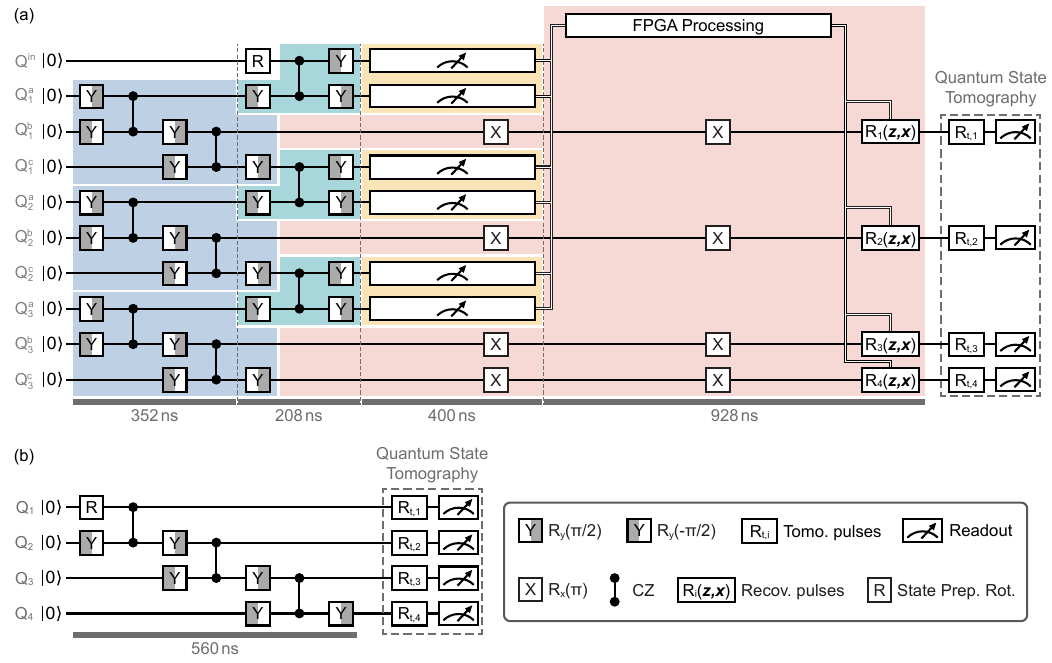}
    \caption{\label{fig:timing_diagram} (a) Decomposing the 1-to-4 fan-out gate sequence with single-qubit rotations and CZ gates. The time budget for each step is shown at the bottom of the figure. Background colors correspond marks the prepare (blue), entangle (cyan), measure (yellow), and feedforward (red) steps depicted in Fig.~\ref{fig:circuit_diagram}. (b) Gate sequence of a unitary 1-to-4 fan out gate.}
\end{figure*}

We realize the 1-to-4 constant-depth fan-out circuit with the native gate set in our architecture, see Fig.~\ref{fig:timing_diagram}. We implement single-qubit x- and y-rotations with 32-ns DRAG pulses~\cite{Motzoi2009}. We realize single-qubit z-rotations with virtual Z-gates~\cite{McKay2017}, with the exception of the feedforward Z gate being decomposed into physical x- and y-rotations $R_x(\pi/2)R_y(\pi)R_x(-\pi/2)$. Further upgrades of the control software would allow for conditional virtual Z-gates and thus reduce this overhead. We decompose CNOT gates into CZ gates~\cite{Rol2019, Negirneac2021} and single-qubit y-rotations. During the feedforward idling time, we apply two Carr-Purcell-Meiboom-Gill (CPMG) dynamical decoupling pulses on every output qubit to mitigate low-frequency noise.

With the aforementioned decomposition, the prepare, entangle, measurement, and feedforward step take 352~ns, 208~ns, 400~ns, and 928~ns, respectively. The whole sequence lasts 1888~ns, with $49\%$ of the time required for the feedforward control. Comparably, the unitary fan-out sequence takes 560~ns due to a shorter circuit depth, see Fig.~\ref{fig:timing_diagram}b.

\section{Error Scaling of the Fan-Out Circuits} \label{app:error_scaling}
Given the error rate of elementary operations, we theoretically analyze the output state fidelity of different fan-out sequences based on a success-probability model. First, we assume that the total error of the sequences $\varepsilon^\mathrm{tot}$ is dominated by the individual CNOT gate errors $\{\varepsilon^\mathrm{CNOT}_i\}$, measurement errors $\{\varepsilon^\mathrm{meas}_j\}$, and idling errors $\{\varepsilon^\mathrm{idle}_k\}$. Then, we assume these errors are incoherent and uncorrelated, which yields an independent success rate of each operation of $(1-\varepsilon)$, where $\varepsilon$ is the error rate of the operation. We estimate the success probability of the full sequence by multiplying the success probability of all elementary operations:
\begin{equation} \label{eq:error_estimation}
1 - \varepsilon^\mathrm{tot} = \prod_{i=1}^{n_\mathrm{CNOT}}(1-\varepsilon^\mathrm{CNOT}_i) \prod_{j=1}^{n_\mathrm{meas}}(1-\varepsilon^\mathrm{meas}_j) \prod_{k=1}^{n_\mathrm{idle}}(1-\varepsilon^\mathrm{idle}_k)\,,
\end{equation}
where $n_\mathrm{idle}$, $n_\mathrm{CNOT}$, $n_\mathrm{meas}$ are the number of error-prone operations of that type. We determine $\{\varepsilon^\mathrm{CNOT}_i\}$ as the sum of one CZ gate error and two single-qubit errors extracted from randomized benchmarking. We obtain $\{\varepsilon^\mathrm{meas}_j\}$ by characterizing single-shot readout (see App.~\ref{app:sample}). We assume that $\varepsilon^\mathrm{idle}_k$ arises from a depolarizing channel acting on each qubit at their Hahn-echo dephasing rate $1/T_2^\mathrm{echo}$ for a duration of the corresponding idling time. 

Equation~\ref{eq:error_estimation} can be used for predicting the error of the sequences, for which all elementary operations have been benchmarked. To predict the error of a sequence at a circuit size beyond the experiment, we replace individual errors with the average error rate of the same type of error source,
\begin{equation} 
1 - \varepsilon^\mathrm{tot} = (1-\bar\varepsilon^\mathrm{CNOT})^{n_\mathrm{CNOT}} (1-\bar\varepsilon^\mathrm{meas})^{n_\mathrm{meas}} (1-\bar\varepsilon^\mathrm{idle})^{n_\mathrm{idle}}\,,
\end{equation}
where $\bar\varepsilon^\mathrm{CNOT},\bar\varepsilon^\mathrm{meas},\bar\varepsilon^\mathrm{idle}$ are the average CNOT, measurement, idling errors for the qubits on the device. For simplicity, we specify $\bar\varepsilon^\mathrm{idle}$ as the idling error of a qubit with a median Hahn-echo dephasing time of all qubits idled for a CNOT gate time $t_\mathrm{CNOT}$. Correspondingly, the occurrence $n_\mathrm{idle}$ is the total idle time of all qubits normalized by $t_\mathrm{CNOT}$. We assume that the feedforward latency introduced by classical electronics is $\mu$ times $t_\mathrm{CNOT}$. In this work, $\mu=7.5$. We list the occurrence of each error source for the three circuits in Table \ref{tab:error_budget_count}.
\begin{table}[ht]
\centering
\begin{tabular}{l>{\centering\arraybackslash}p{2cm}>{\centering\arraybackslash}p{1.5cm}>{\centering\arraybackslash}p{1.5cm}}
\toprule
Circuit &  $n_\mathrm{idle}$ &  $n_\mathrm{CNOT}$ & $n_\mathrm{meas}$ \\
\midrule
Unitary & $(n^2-3n+2)/2$  & $n-1$  & $0$ \\
Feedforward & $n(\mu +2)-1$ & $3n-3$ & $2n-2$ \\
Pauli frame update  & $2n-1$         & $3n-3$ & $2n-2$ \\
\bottomrule
\end{tabular}
\caption{\label{tab:error_budget_count} Occurrences $n_\mathrm{idle}$,  $n_\mathrm{CNOT}$, and $n_\mathrm{meas}$ of the dominating error sources for the three fan-out circuits discussed in the main text, vs.~the number of output qubits $n$.}
\end{table}

We note that in the limit of small output qubit number $n$, $\varepsilon^\mathrm{tot}$ scales quadratically with $n$ for the unitary protocol and linearly with $n$ for the constant-depth protocols (i.e.~the feedforward and Pauli frame update implementations).
\bibliography{FanOutGateRef}
\end{document}